# A distorted-wave approach to the elastic scattering of twisted electrons

A. L. Harris[1] and S. Fritzsche[2,3,4]

[1]Department of Physics, Illinois State University, Normal, IL 61790, USA
[2]Helmholtz-Institut Jena, 07743 Jena, Germany
[3]Theoretisch-Physikalisches Institut, Friedrich-Schiller-Universität Jena, 07743 Jena, Germany
[4]GSI Helmholtzzentrum für Schwerionenforschung GmbH, 64291 Darmstadt, Germany

**Abstract**

The elastic scattering of spinless vortex electrons on realistic target atoms has been investigated. In particular, expressions are derived in different approximations for the elastic angular-differential cross sections. We develop a distorted wave formalism that includes the effect of the atomic potential on the impinging vortex electron and compare this to a plane-wave Born approximation without such a distortion. Detailed computations have been performed for elastic scattering of vortex electrons on helium, neon, and argon targets by varying the energy, topological charge, and opening angle. Our results show that the overall magnitude of the cross section increases when the distortion by the bound-state electrons is taken into account. We also show that under certain conditions, such as high-Z targets or projectiles with low values of topological charge, significant differences in cross section shape and magnitude are observed between the distorted-wave and plane-wave Born models. Thus, the plane-wave Born approximation must be used with caution when describing vortex electron collisions.

## I. Introduction

The scattering of an electron from a potential is one of the most fundamental atomic physics processes. As such, plane-wave (or non-vortex) electron scattering has been considered for many decades, and an excellent understanding of the physical interactions involved has been achieved. In recent years, there have been a growing number of investigations into non-plane-wave (or vortex) electrons. These so-called vortex or twisted electrons are unique in their ability to carry quantized orbital angular momentum (OAM) and their non-zero transverse momentum. They possess a phase vortex with a local phase singularity in the form $e^{i\lambda\phi}$, where $\lambda$ is an integer referred to as the topological charge and indicates the vortex electron's OAM. The impinging electron has a spiral wave front with a node at its center.

Theoretically, there have been many investigations into collisions between vortex electrons and atomic and molecular targets [1–13]. These works have provided valuable initial insight into vortex-electron-induced processes, such as elastic scattering, excitation, and ionization, and the growing body of literature on vortex electron collisions has shown unique features towards new physics applications. For example, because vortex electrons carry quantized OAM, it has been shown that OAM can be exchanged between the twisted electron and both the electronic and center of mass motions of the atomic target [6,7]. This exchange of OAM results in an alteration of the selection rules for excitation processes [8] and OAM transfer to continuum electrons in ionization collisions [14]. The OAM of vortex electrons is also predicted to allow for the observation of dichroism in inelastic scattering from chiral molecules [15] as well as to study [16] and control [13] interference features in cross sections for molecular targets. The non-zero transverse momentum of twisted electrons also alters the angular distribution of ejected electrons in ionization collisions [5,17] and can result in emission of secondary electrons in directions that are otherwise forbidden [12]. Vortex electrons also open the door to studies of fundamental properties of matter, including access to the Coulomb phase through elastic scattering [18], electron projectile coherence properties [13,19], the Faraday effect for electrons in vacuum [20], and the study of forbidden transitions, electron correlations, and relativistic effects through hyperfine structure [21].

To date, most of the existing calculations for twisted electron collisions have relied on simple models such as the Born approximation with single active electron targets or (local) analytic approximations to the scattering potential. For collisions with plane-wave electrons, it is well-known that these approximations are not able to well-describe the collision process. For non-vortex (plane-wave) projectiles, for example, the Born approximation underestimates the magnitude of the cross section and is generally unable to accurately account for higher order collisions and short-range interactions [22]. By their very nature, in addition, single active electron models cannot include multi-electron effects.

Here, we address the shortcomings of existing twisted electron models by developing a distorted wave model and analyzing the elastic scattering of vortex electrons from the realistic scattering potentials of multi-electron targets, such as helium, neon, and argon. The use of a distorted wave model allows for a more accurate description of the twisted electron's interaction with the target and is applicable to model collisions with low-energetic projectiles. These improvements allow us to assess the conditions under which the Born approximation is valid for vortex electron scattering and to accurately calculate cross sections for multi-electron targets at lower projectile energies.

Our results show that the use of a distorted wave approximation increases the magnitude of the cross sections when compared to the plane-wave Born approximation, similar to what is observed in non-vortex collisions. This enhancement is independent of projectile energy and vortex opening angle. The overall shape of the cross section is typically not altered by the use of the distorted wave approximation, except for the scattering of vortex electrons with zero OAM or scattering from an argon target. The largest differences in magnitude between the two approximations appear for small values of OAM and these differences diminish as OAM increases. This indicates that for higher collision energies or tightly bound electrons, the distortion by the target becomes negligible. However, for high-Z targets or vortex projectiles with small values of OAM, atomic distortion effects significantly alter the shape and magnitude of the cross sections. Atomic units are used throughout unless otherwise noted.

**II. Theory**

We first consider a spinless electron elastically scattering from helium which forms a spherically symmetric (local) potential. This potential is generated self-consistently based on the Hartree-Fock-Slater method by including the major exchange contributions for the bound electrons [23]. The exchange interaction between the incident projectile electron and the target is assumed to be negligible and not included in the computations. Such self-consistent calculations predict the low-lying excitation and electron binding energy typically with an accuracy of a few percent.

Our goal is to derive expressions for the elastic scattering cross sections for Bessel electrons. We consider 'head-on' collisions with isolated atoms at the axis of the vortex electron beam (see Fig. 1a).

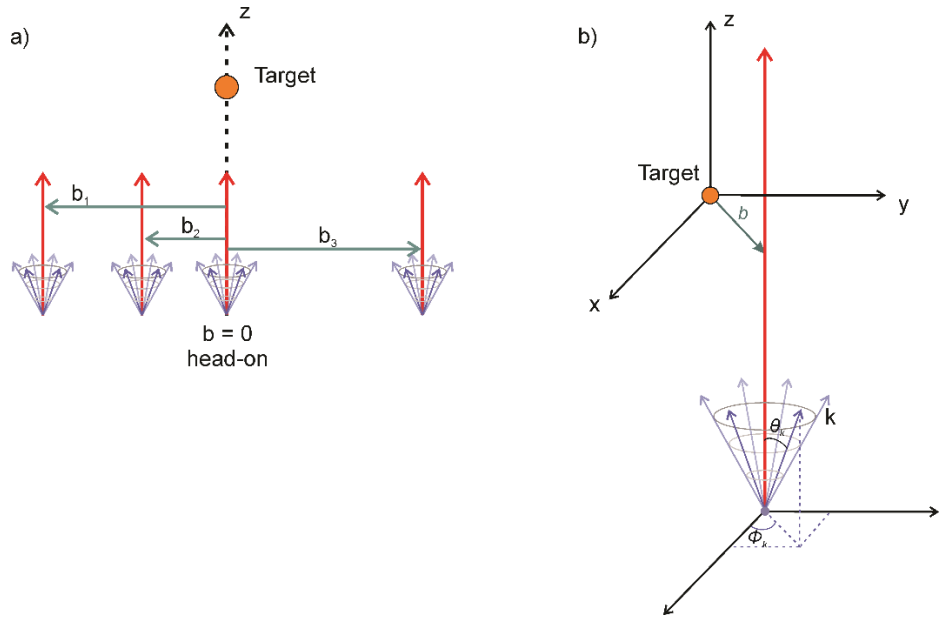

Figure 1 (a) Depiction of different vortex projectile impact parameters $\vec{b}$ (green arrows). The vertical red arrows represent the propagation direction of the incident projectile, and the incident vortex momentum vectors lie on a cone of half angle $\theta_k$ (purple vectors). The 'head-on' geometry of $\vec{b} = 0$ occurs when the atom is aligned with the incident projectile's propagation axis. (b) Collision geometry for elastic scattering with an incident vortex projectile. As in (a), the projectile propagates along the direction of the red arrow with an impact parameter of $\vec{b}$. Each momentum vector has an azimuthal angle of $\phi_k$ and a vortex opening angle of $\theta_k$.

## A. Distorted wave approach

The electron wave function $\psi(\vec{r})$ is a solution to the Schrödinger equation

$$[\nabla^2 - U(r) + k^2]\psi(\vec{r}, k) = 0 \tag{1}$$

where $k^2 = 2E$, $U(r) = 2V(r)$ in atomic units, $\vec{k}$ is the wave vector with polar and azimuthal coordinates $(\theta_k, \phi_k)$, and $E$ is the energy of the electron. Following the standard distorted wave formalism [24], in spherical coordinates, the wave function can be written in terms of a partial wave expansion

$$\psi(\vec{r}, k) = \frac{1}{kr} \sum_{l=0}^{\infty} \sum_{m=-l}^{l} u_{lm}(r, k) Y_{lm}(\theta, \phi), \tag{2}$$

where each term is itself a solution to the Schrödinger equation and $(\theta, \phi)$ are the polar and azimuthal coordinates of $\vec{r}$. The radial functions $u_{lm}(r, k)$ satisfy the radial Schrödinger equation

$$\frac{d^2 u_{lm}(r,k)}{dr^2} + \left[k^2 - U(r) - \frac{l(l+1)}{r^2}\right] u_{lm}(r, k) = 0. \tag{3}$$

and $Y_{lm}(\theta, \phi)$ is a spherical harmonic function. The quantum numbers $l$ and $m$ are the orbital angular momentum (eigenfunction of $L^2$) and magnetic quantum numbers (eigenfunction of $L_z$), respectively. The radial functions $u_{lm}(r, k)$ are normalized such that the wave function (Eq. (2)) takes asymptotically the form

$$\psi(\vec{r}, k) \to \psi_{free} + \frac{e^{ikr}}{r} f(\theta, \phi), \tag{4}$$

a sum of free-particle motion (without any potential) and a spherical wave with scattering amplitude $f(\theta, \phi)$. This amplitude depends on the polar and azimuthal angles and is used to calculate the cross section

$$\frac{d^2\sigma}{d\Omega} = |f(\theta, \phi)|^2. \tag{5}$$

Asymptotically, for a finite range potential, Eq. (3) becomes

$$\frac{d^2 w_{lm}(r,k)}{dr^2} + \left[k^2 - \frac{l(l+1)}{r^2}\right] w_{lm}(r, k) = 0, \tag{6}$$

where $w_{lm}(r,k) = u_{lm}(r \to \infty, k)$ is the asymptotic form of the radial function given by

$$w_{lm}(r,k) = N_{lm} \sin\left(kr - \frac{l\pi}{2} + \delta_l\right). \tag{7}$$

The normalization constant is $N_{lm}$ and $\delta_l$ is the partial wave phase shift. To proceed further and determine the normalization constant and scattering amplitude, it is necessary to specify the free particle wave function $\psi_{free}$, which in our case can be either a (non-vortex) plane wave or a vortex Bessel wave. We begin first with the familiar non-vortex case and then show the analogous derivation for the vortex wave.

### B. Non-vortex electrons

In this case, the free particle wave function is given by a plane wave

$$\psi_{free} = e^{i\vec{k}\cdot\vec{r}}. \tag{8}$$

For the most general propagation direction $\vec{k}$, the normalization constant and scattering amplitude is obtained by inserting Eq. (8) into Eq. (4) and equating to Eq. (2) using the asymptotic form of the radial function from Eq. (7). This yields the following

$$e^{i\vec{k}\cdot\vec{r}} + \frac{e^{ikr}}{r} f^{(NV)}(\theta,\phi) = \frac{1}{kr} \sum_{l=0}^{\infty} \sum_{m=-l}^{l} N_{lm}^{(NV)} \sin\left(kr - \frac{l\pi}{2} + \delta_l\right) Y_{lm}(\theta,\phi). \tag{9}$$

The partial wave expansion for a plane wave propagating along a general propagation direction is given by [24]

$$e^{i\vec{k}\cdot\vec{r}} = \frac{4\pi}{kr} \sum_{l=0}^{\infty} \sum_{m=-l}^{l} i^l F_l(kr) Y_{lm}^*(\theta_k, \phi_k) Y_{lm}(\theta,\phi), \tag{10}$$

where $F_l(kr)$ is the regular spherical Bessel function that has the asymptotic form

$F_l(kr) \xrightarrow[r\to\infty]{} \sin\left(kr - \frac{l\pi}{2}\right)$. By combining Eqs. (9) and (10), the following expressions for the normalization constant and scattering amplitude for a non-vortex (NV) projectile are found

$$N_{l,m}^{(NV)} = e^{i\delta_l} 4\pi i^l Y_{lm}^*(\theta_k, \phi_k) \tag{11}$$

and

$$f^{(NV)}(\theta,\phi) = \frac{1}{k}\sum_{l=0}^{\infty}\sum_{m=-l}^{l} e^{i\delta_l} 4\pi\, Y_{lm}^*(\theta_k,\phi_k) Y_{lm}(\theta,\phi) \sin\delta_l. \tag{12}$$

From the expressions (11-12), we obtain the familiar form from textbooks for the propagation along the z-axis with $m = 0$ and $\theta_k = 0$ as

$$N_{l,m=0}^{(NV)} = e^{i\delta_l} 4\pi i^l \sqrt{\frac{(2l+1)}{4\pi}} \tag{13}$$

$$f_{m=0}^{(NV)}(\theta; m=0) = \frac{1}{k}\sum_{l=0}^{\infty}(2l+1)e^{i\delta_l} \sin\delta_l\, P_l(\cos\theta). \tag{14}$$

Physically, the phase shifts $\delta_l$ contain the information about the atomic potential's distortion on the projectile electron. These shifts reflect the strength and range of the incident electron interacting with the atom. For each partial wave, the phase shift is obtained by matching the free-electron solution of the Schrödinger equation with those for moving in the atomic potential [25]. Apart from the cross sections, these phase shifts also give rise to the scattering length or the spin-polarization of the scattered electrons.

## C. Vortex electrons

For vortex electrons, we consider a Bessel wave, although other electron vortex waveforms exist. The Bessel electronic wave function is also a solution to the free particle Schrödinger equation, but is expressed most conveniently in cylindrical coordinates $(\rho, \varphi, z)$

$$\psi_{free} = \psi_{Bess}(\vec{r}, k_\rho, k_z, \lambda) = A_\lambda J_\lambda(k_\rho \rho) e^{i\lambda\varphi} e^{ik_z z} e^{i\vec{k}\cdot\vec{b}}, \tag{15}$$

where $\lambda$ is the topological charge equal to the electron's orbital angular momentum, $A_\lambda$ is a normalization constant, $J_\lambda(k_\rho\rho)$ is a Bessel function, and $\vec{k} = (k_\rho, \phi_k, k_z)$ is the projectile's momentum. An important distinction between the plane-wave electron and the Bessel electron is that the Bessel electron has both transverse and longitudinal components of momentum and its probability density is not spatially uniform in the transverse plane. Therefore, the displacement of the Bessel electron's transverse center relative to the atom must be specified. This transverse displacement is referred to as the impact parameter $\vec{b}$ (see Fig. 1).

In addition to Eq. (15), the Bessel wave function can be conveniently expressed as a linear combination of plane waves with each plane wave momentum $\vec{k}$ lying on a cone of half angle $\theta_k$.

$$\psi_{Bess}(\vec{r}, k_\rho, k_z, \lambda) = \frac{1}{(2\pi)} \int d^2 k_\perp a_{k_\rho,\lambda}(\vec{k}_\perp) e^{i\vec{k}\cdot\vec{r}} e^{i\vec{k}_\perp \cdot \vec{b}}, \qquad (16)$$

where $a_{k_\rho,\lambda}(\vec{k}_\perp) = (-i)^\lambda e^{i\lambda\phi_k} \frac{\delta(k_\perp - k_\rho)}{k_\rho}$. The half angle of the cone is referred to as the opening angle and relates the transverse and longitudinal momenta $k_\rho$ and $k_z$

$$\tan\theta_k = k_\rho/k_z. \qquad (17)$$

The opening angle is experimentally controllable and is a fixed value for a given collision process.

Following a similar process as in Section B for a non-vortex electron, the normalization constant and scattering amplitude can be found for the Bessel vortex electron. If the wave function $\psi_{Bess}(\vec{r}, k_\rho, k_z, \lambda)$ of Eq. (16) is inserted into the asymptotic wave function expression of Eq. (4) and equated to Eq. (2) using the asymptotic form of the radial function from Eq. (7), we obtain the following expression for a vortex electron analogous to the non-vortex expression of Eq. (9)

$$\frac{1}{(2\pi)} \int d^2 k_\perp a_{k_\rho,\lambda}(\vec{k}_\perp) e^{i\vec{k}\cdot\vec{r}} e^{i\vec{k}_\perp\cdot\vec{b}} + \frac{e^{ikr}}{r} f^{(Bessel)}(\theta, \phi, \theta_k, \lambda, \vec{b}) = \frac{1}{kr} \sum_{l=0}^{\infty} \sum_{m=-l}^{l} N_{lm}(\theta_k, \lambda, \vec{b}) \sin\left(kr - \frac{l\pi}{2} + \delta_l\right) Y_{lm}(\theta, \phi), \quad (18)$$

where the scattering amplitude $f^{(Bessel)}(\theta, \phi, \theta_k, \lambda, \vec{b})$ for the Bessel electron now additionally depends on the opening angle, topological charge, and impact parameter.

Again, using the partial wave expansion for a plane wave from Eq. (10) yields the key expressions for the normalization constant and scattering amplitude for a Bessel electron within the distorted wave formalism

$$N_{lm}^{(Bessel)}(\theta_k, \lambda, \vec{b}) = 2e^{i\delta_l} \int d^2 k_\perp a_{k_\rho,\lambda}(\vec{k}_\perp) e^{-i\vec{k}_\perp\cdot\vec{b}} i^l Y_{lm}^*(\theta_k, \phi_k) \qquad (19)$$

and

$$f^{(Bessel)}(\theta,\phi,\theta_k,\lambda,\vec{b}) = \frac{1}{k}\sum_{l=0}^{\infty}\sum_{m=-l}^{l} N_{lm}^{(Bessel)}(\theta_k,\lambda,\vec{b}) Y_{lm}(\theta,\phi)\, i^{-l} \sin\delta_l. \qquad (20)$$

Note that the expression for the scattering amplitudes for the vortex (Eq. (20)) and non-vortex electrons (Eq. (12)) are identical, except for the difference in normalization constants.

By inserting the normalization constant from Eq. (19) and the Fourier expansion coefficient $a_{k_\rho,\lambda}(\vec{k}_\perp)$ into Eq. (20), the Bessel scattering amplitude can be written in terms of the non-vortex scattering amplitude

$$f^{(Bessel)}(\theta,\phi,\theta_k,\lambda,\vec{b}) = (-i)^\lambda \int d\phi_k\, e^{i\lambda\phi_k} e^{-i\vec{k}_\perp\cdot\vec{b}} f^{(NV)}(\theta,\phi). \qquad (21)$$

**i. Head-on collisions**

The non-uniform spatial profile of the vortex electron requires the distinction of different collision geometries that depend on the impact parameter. For a head-on collision, the target atom is aligned with the transverse center of the vortex wave, corresponding to an impact parameter of $\vec{b} = 0$. In this case, the normalization constant and scattering amplitude become

$$N_{l,m}^{(Bessel)}(\theta_k,\lambda,\vec{b}=0) = 2e^{i\delta_l}(-i)^\lambda i^l \int d\phi_k\, e^{i\lambda\phi_k} Y_{lm}^*(\theta_k,\phi_k) = 4\pi e^{i\delta_l}(-i)^\lambda i^l (-1)^m \left[\frac{(2l+1)(l-m)!}{(4\pi)(l+m)!}\right]^{\frac{1}{2}} P_l^m(\cos\theta_k)\delta_{\lambda,m} \qquad (22)$$

and

$$f^{(Bessel)}(\theta,\phi,\theta_k,\lambda,\vec{b}=0) = (-i)^\lambda \int d\phi_k\, e^{i\lambda\phi_k} f^{(NV)}(\theta,\phi) = \frac{4\pi(-i)^\lambda}{k}\sum_{|\lambda|\le l}^{\infty} e^{i\delta_l}(-1)^\lambda \left[\frac{(2l+1)(l-\lambda)!}{(4\pi)(l+\lambda)!}\right]^{\frac{1}{2}} P_l^\lambda(\cos\theta_k) Y_{l\lambda}(\theta,\phi)\sin\delta_l. \qquad (23)$$

Equations (22) and (23) are used to calculate the vortex distorted wave cross sections shown below in Figs. 2-5 ('vDWA'; solid black curves). Note that in the limit that $\lambda = 0$ and $\theta_k = 0$, the non-vortex expressions are recovered from Eqs. (22) and (23).

**C. Plane-Wave Born Approximation**

The distorted wave approach taken above includes the atomic potential in distorting the projectile wave function. If this distortion is neglected, one recovers what is commonly referred to as the plane-wave Born approximation (PWBA), which, until now, has been the primarily applied treatment for vortex collisions. In the PWBA, the radial functions are found by setting the distorting potential $U(r) = 0$ in Eq. (3) and are given by

$$u_{lm}(r,k) = F_l(kr) = krj_l(kr)), \tag{24}$$

where $j_l(kr))$ is a spherical Bessel function. These radial functions are then used to calculate the phase shifts for a non-zero potential

$$\tan \delta_l^{Born} = -\frac{1}{k}\int_0^\infty dr\, F_l^2(kr)U(r). \tag{25}$$

Using the Born phase shifts $\delta_l^{Born}$ in Eqs. (22) and (23) leads to the scattering amplitude and cross section within the vortex PWBA and is designated by vPWBA (red dashed curves) in Figs. 2-5.

### III. Results and Discussion

To examine the validity of the PWBA for vortex electron collisions, we present cross section calculations for the vPWBA and vDWA models for a number of different vortex parameters. Whereas the projectile energy is typically the primary collision parameter for non-vortex projectiles, additional degrees of freedom can be controlled with vortex electrons. In addition to the projectile energy, vortex electrons have quantized OAM and transverse momentum. These properties can be selected to aid in the control of the dynamics or outcome of the collision process. We investigate the effect of each of these vortex properties on the vDWA elastic scattering cross section when multi-electron effects and the influence of the atomic potential are included in the the calculated angular differential cross sections.

Figure 2a-d shows a comparison of vPWBA and vDWA cross sections for a head-on collision of a 20 eV vortex electron with an opening angle of $\theta_k = 15°$. Calculations were performed for OAM values between $\lambda = 0$ and 5. For $\lambda = 0$, both the vPWBA and the vDWA cross sections exhibit a maximum at $\theta_s = 0°$ and a sharp decrease in magnitude as $\theta_s$ increases, similar to what has been observed in the non-vortex cross sections. In this case, the inclusion of the atomic potential into the vDWA calculations increases the backward scattering by more than an order of magnitude and results in a nearly isotropic scattering distribution between $\theta_s = 90°$ and $\theta_s = 180°$. However, for $\lambda > 0$, both the vPWBA and vDWA cross sections show a zero in the forward direction at $\theta_s = 0°$, with a peak at small $\theta_s$, and a significant decrease in magnitude for larger scattering angles. The zero in the vortex cross section has been previously reported in vPWBA [9] calculations for elastic scattering from a Yukawa potential. Its origin was traced to the zero intensity at the center of the vortex wave's spatial density and is indicative of emission of a vortex wave from the scattering center [9]. In the case of $\lambda = 0$, no node is present in the vortex wave's transverse density (see Fig. 2e) and thus no zero appears in the cross section. Our distorted wave calculations show that this feature persists even when the atomic potential is taken into account.

In general, the magnitude of the vDWA cross sections are greater than the vPWBA cross sections, similar to what has been observed with non-vortex cross sections [22]. This magnitude difference is greatest for projectiles with small values of OAM, while for $\lambda = 5$, there is very little difference between the vDWA and vPWBA cross sections. The similarity of the vDWA and vPWBA cross sections for larger OAM values is likely due to the overall reduction in the cross section magnitude as OAM increases. As Fig. 2f-h shows, the node in the vortex wave transverse profile becomes wider as the OAM increases, resulting in less overlap between the projectile and the scattering potential. Additionally, the magnitude of the vortex transverse density decreases with larger OAM. Combined, these two features result in an overall reduction in magnitude of the scattering cross section for larger OAM values for both the vPWBA and vDWA models. At the largest value of OAM ($\lambda = 5$), there is very little overlap between the projectile wave and the scattering center causing any atomic distortion effects included in the vDWA calculation to be negligible.

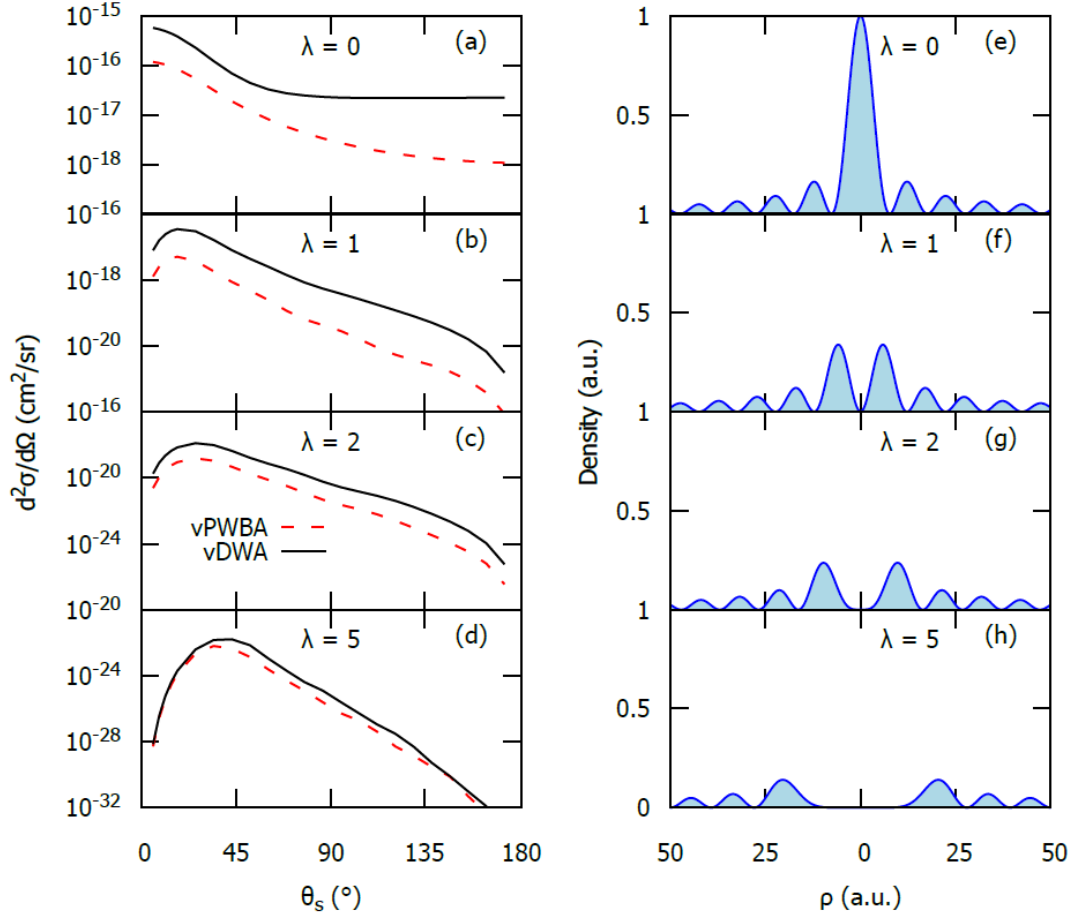

Figure 2 (a-d) Angular differential cross sections $\frac{d^2\sigma}{d\Omega}$ as a function of projectile scattering angle $\theta_s$ for electron-impact elastic scattering from helium. The incident projectile energy is 20 eV and the vortex opening angle is $\theta_k = 15°$. Results are shown for head-on collisions using a vortex Bessel electron in the vortex plane-wave Born approximation (vPWBA, dashed red line) and with the vortex distorted wave model (vDWA, solid black line) for topological charge of (a) $\lambda = 0$, (b) $\lambda = 1$, (c) $\lambda = 2$, (d) $\lambda = 5$. (e-h) Bessel wave function density in the transverse plane as a function of radial distance for the Bessel waves used in (a-d).

Figure 3a-c shows a comparison of the vPWBA and vDWA cross sections for a head-on collision of a 20 eV vortex electron with an OAM value of $\lambda = 2$. Calculations were performed for opening angles between $\theta_k = 5°$ and $30°$. As expected, because the OAM is non-zero, the cross sections exhibit a zero in the forward direction. The peak in the cross sections broadens and shifts to larger scattering angles as the opening angle increases, and its location occurs at approximately $\theta_s = \theta_k$. Again, the vDWA cross section is larger than the vPWBA cross section, although this difference in magnitude does not change with the opening angle, indicating that the effects of the atomic potential are not dependent on the projectile's transverse momentum ($k_\rho = k \sin \theta_k$). For both the vDWA and vPWBA approximations, the overall magnitude of the cross section increases for larger opening angles. This increase of the cross section can again be traced to the vortex projectile's transverse density. Fig. 3d-f shows that the transverse density of the vortex projectile is approximately the same magnitude, regardless of opening angle, but the central node is more narrow for larger opening angles. This results in greater overlap between the projectile wave packet and the scattering potential for projectiles with large opening angles and results in larger cross sections. This increased overlap occurs regardless of whether atomic distortion effects are included.

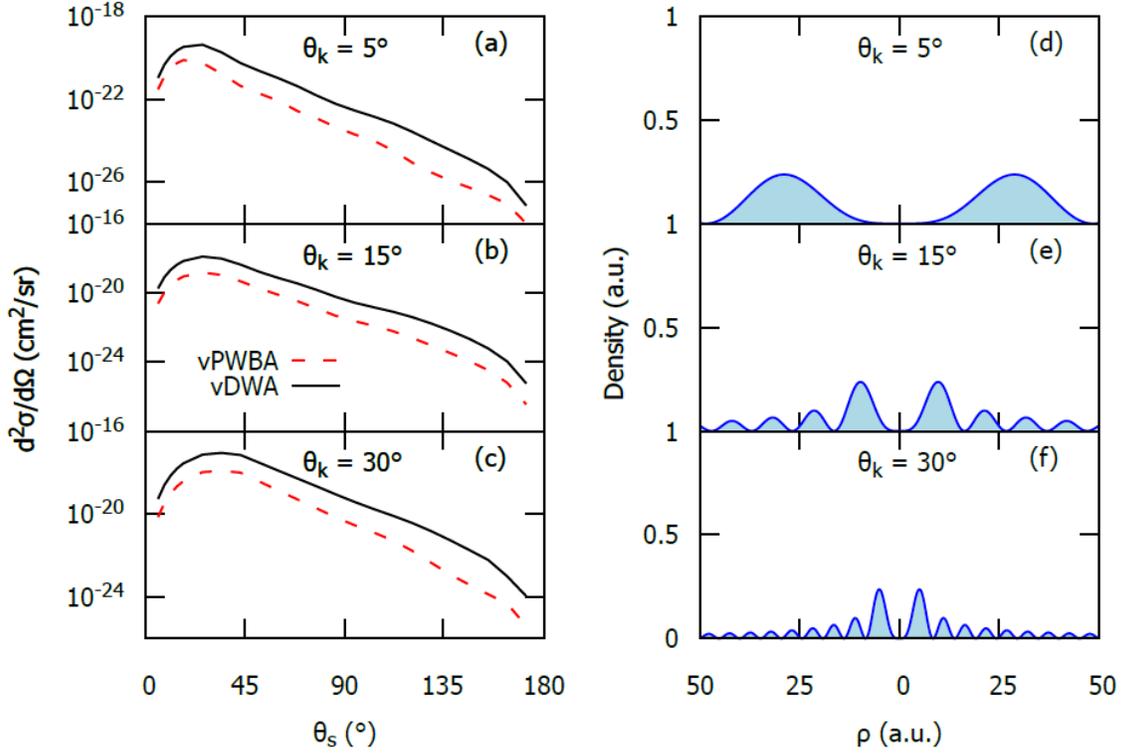

Figure 3 (a-c) Angular differential cross sections $\frac{d^2\sigma}{d\Omega}$ as a function of projectile scattering angle $\theta_s$ for electron-impact elastic scattering from helium. The incident projectile energy is 20 eV and the topological charge is $\lambda = 2$. Results are shown for head-on collisions using a vortex electron in the vortex plane-wave Born approximation (vPWAB, dashed red line) and with the vortex distorted wave model (vDWA, solid black line) for opening angles of (a) $\theta_k = 5°$, (b) $\theta_k = 15°$, (c) $\theta_k = 30°$. (d-f) Bessel wave function density in the transverse plane as a function of radial distance for the Bessel waves used in (a-c).

Figure 4a-c shows a comparison of the vPWBA and vDWA cross sections for a head-on collision of a Bessel electron with an OAM value of $\lambda = 2$ and opening angle of $\theta_k = 15°$. Cross sections are shown for three projectile energies, $E_i = 10$, 20 and 50 eV. The cross sections are again zero in the forward direction due to the projectile's non-zero OAM and the vDWA cross section is larger in magnitude than the vPWBA cross section. The difference in magnitude between the two approximations remains fairly consistent as projectile energy changes and the magnitude of the cross sections is only slightly affected by the projectile energy. For larger

projectile energies, the peak in the cross sections becomes narrower and shifts to smaller scattering angles. This indicates that higher energy projectiles scatter at smaller scattering angles, consistent with observations from non-vortex collisions.

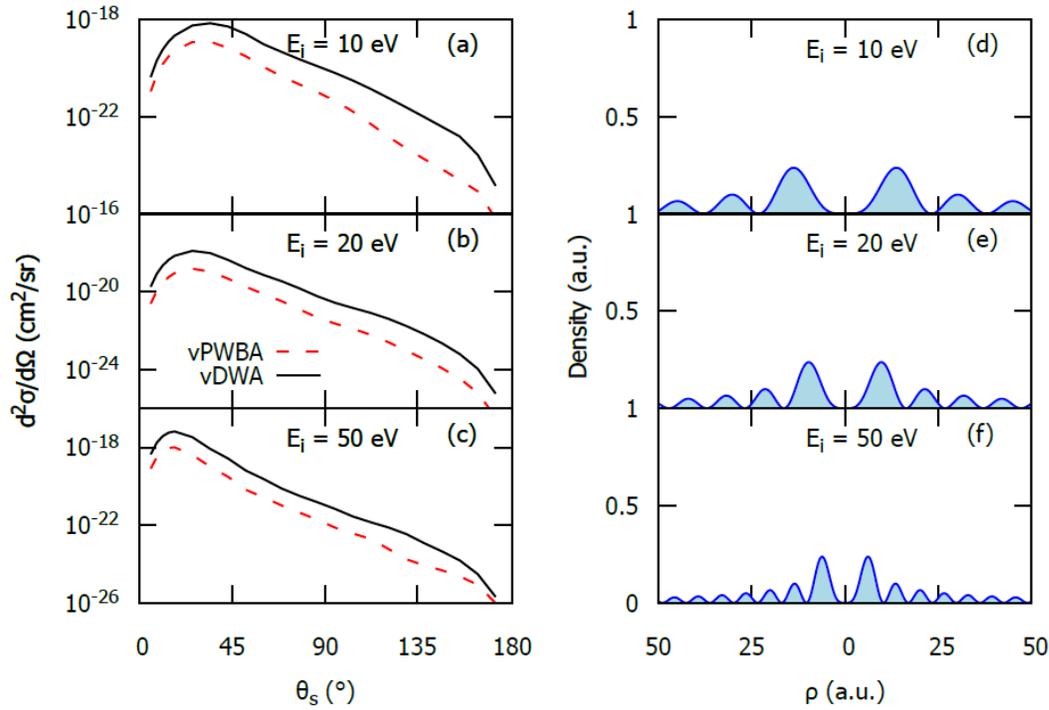

Figure 4 (a-b) Angular differential cross sections $\frac{d^2\sigma}{d\Omega}$ as a function of projectile scattering angle $\theta_s$ for electron-impact elastic scattering from helium. The vortex projectile opening angle is $\theta_k = 15°$ and the topological charge is $\lambda = 2$. Results are shown for head-on collisions using a vortex electron in the vortex plane-wave Born approximation (vPWBA, dashed red line) and with the vortex distorted wave model (vDWA, solid black line) for energies of (a) $E = 10$ eV, (b) $E = 20$ eV, (c) $E = 50$ eV. (d-f) Bessel wave function density in the transverse plane as a function of radial distance for the Bessel waves used in (a-c).

Let us finally examine how the use of a distorted wave model affects the cross sections for realistic atomic potentials. To this end, Fig. 5a-c compares the vPWBA and vDWA cross sections for helium, neon, and argon targets. The vPWBA cross sections are similar in shape and magnitude for all three targets, but the argon vDWA cross sections show a pronounced backward peak. This enhanced backward scattering is only present in the vDWA calculation and likely reflects the shell-structure of complex atoms. Apart from the individual contributions of the (bound) L- and M-shell electrons to the overall charge distribution, this shell structure may cause interferences in the scattering amplitude. It will be interesting to further explore this departure of the scattering cross sections when compared to the short-range potential of atomic helium. The dramatic difference in vPWBA and vDWA cross sections for argon indicate that atomic distortion effects from realistic potentials cannot be neglected, particularly for high-Z targets.

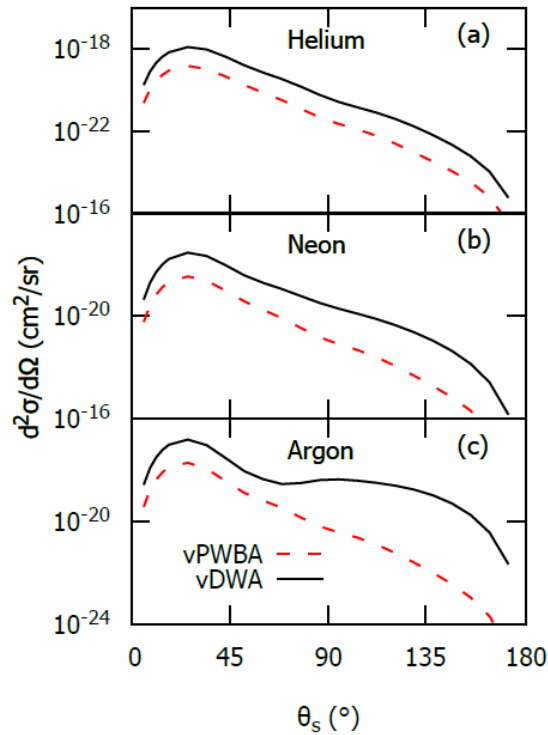

Figure 5 (a-b) Angular differential cross sections $\frac{d^2\sigma}{d\Omega}$ as a function of projectile scattering angle $\theta_s$ for electron-impact elastic scattering from helium, neon, and argon. The vortex projectile has an energy of 20 eV, an opening angle of $\theta_k = 15°$, and a topological charge of $\lambda = 2$. Results are shown for vortex electrons in the vortex plane-wave Born approximation (vPWBA, dashed red line) and the vortex distorted wave model (vDWA, solid black line).

### IV. Conclusions

We have developed a new distorted wave formalism for the evaluation of vortex-electron-impact elastic scattering cross sections from realistic atomic potentials. This formalism allows for the inclusion of the (detailed) atomic potential to the scattering of electrons. We present cross sections using our vDWA model for elastic scattering from helium, neon, and argon across a range of physically-tunable projectile parameters, such as topological charge, opening angle, and energy. These cross sections are compared to cross sections calculated within the vPWBA in order to determine the effects of atomic distortion.

Our results show that cross sections calculated using the vDWA model are larger in magnitude than those calculated with the vPWBA model. This is consistent with similar trends seen in previous works on non-vortex electrons. As the projectile's topological charge increases, the effect of atomic distortion is reduced and the vDWA and vPWBA cross sections become more similar. This effect is traced to the reduced overlap between the projectile wave packet and the target potential, resulting in atomic distortion effects having less influence on the scattering process. The difference in magnitude between the vPWBA and vDWA cross sections is not significantly affected by the vortex projectile's opening angle or projectile energy.

For an argon target, the vDWA cross sections exhibit a strong backward scattering peak that is not present in the cross sections calculated with the vPWBA model or for other targets. This backward peak is likely caused by L- and M-shell effects in the argon scattering potential and indicates that the use of a realistic scattering potential and atomic distortion effects can significantly alter the shape and magnitude of the cross sections.

Overall, our results demonstrate that under certain conditions, such as low-Z targets and high topological charge, the atomic potential's effect on the electron is minimal. However, these effects can significantly alter the cross sections in both shape and magnitude under other conditions and may be required to accurately describe vortex electron scattering.

## V. Acknowledgements

A. L. H. acknowledges the support of the National Science Foundation under Grant No. PHY-2207209.